# Induced magnetic ordering in alloyed compounds based on Pauli paramagnet $YCo_2$

Z. Śniadecki, M. Werwiński, A. Szajek

Institute of Molecular Physics, Polish Academy of Sciences

M. Smoluchowskiego 17, PL-60-179 Poznań, Poland

U.K. Rößler

IFW Dresden, P.O. Box 270116, D-01171 Dresden, Germany

B. Idzikowski

Institute of Molecular Physics, Polish Academy of Sciences

M. Smoluchowskiego 17, PL-60-179 Poznań, Poland

**Abstract**

Intermetallic $YCo_2$ compound is a Pauli exchange-enhanced paramagnet. Structural and magnetic properties of rapidly quenched $YCo_2$ and $YCo_2$ alloyed with Nb or Ti are presented. Samples produced by melt spinning have been characterized by X-ray diffraction (XRD) and vibrating sample magnetometry (VSM). The samples crystallize in $MgCu_2$-type phase with lattice constant *a* changing from 7.223 Å for $YCo_2$, through 7.213 Å for $Y_{0.9}Nb_{0.1}Co_2$ to 7.192 Å for $Y_{0.9}Ti_{0.1}Co_2$, where Y atoms are replaced by Nb or Ti atoms. Nanocrystalline phases can be produced by appropriate cooling rates for the solidification process. By the synthesis process free volumes, vacancies, and alloyed atoms are introduced into the $YCo_2$ intermetallic. *Ab-initio* calculations have been performed to investigate the effects of substitution on the spin-split electronic band structure in the ordered $YCo_2$. A ferrimagnetic ground state is found in the alloyed systems with substitution on the Y-site which is energetically favorable compared to point defects on Co-sites. However, the experimentally found increased magnetic ordering in alloyed $YCo_2$ appears to be based on microstructure effects.

KEYWORDS: Pauli paramagnet, *ab-initio* calculation, itinerant magnetism, nanocrystalline structure

**Introduction**

The Y-Co compounds have characteristic profile of the electronic density of states (DOS) curve near the Fermi level[1] which can be described as the association of narrow *3d* band of Co with wider *4d* band of Y. These two bands are close to each other and strong hybridization can occur between them. By alloying *3d* element (Co) with nonmagnetic rare-earth metal (Y), the *3d* band becomes wider, the DOS at the Fermi level decreases, and ferromagnetic properties weaken with the increasing Y concentration.[2] The variations of the magnetic moment value can also be explained on the basis of a charge transfer whereby electrons are relocated from less electronegative Y to *3d* component.

In the Y-Co family the Laves phase compound $YCo_2$ with the cubic C15 structure is an interesting example for surface- or defect-related magnetism. Owing to a strong sensitivity of the magnetic ordering on structure and/or microstructure and thermodynamic control parameters, this compound displays various unusual effects, like surface magnetic ordering in films occurring at higher temperatures than the bulk Curie point,[3,4] and enhanced ferromagnetism in nanocrystalline alloys.[2]

On the other hand, the proximity of ferromagnetic order and its sensitive dependence both on thermodynamic control parameters and on details in the real structure of such materials pose a number of fundamental problems in magnetism. *E.g.*, crystalline $YCo_2$ has been investigated because of its itinerant-electron metamagnetism (IEM).[5,6] This field-driven first-order transition from the paramagnetic ground-state to a ferromagnetic state relates to a change in the band structure of *3d* electrons by applying a magnetic field.

In contrast to the paramagnetic behavior in the Laves phase structure, the amorphous counterpart is a ferromagnet.[7] Therefore, subtle differences in the *3d* band structure can influence the IEM behavior which causes a remarkable dependence of magnetism on lattice defects in $YCo_2$. Ferromagnetism with high Curie temperature has also been found in amorphous fine particles of $YCo_2$.[8]

In order to assess the relevance of microstructural features in nanocrystalline alloys for their outstanding magnetic properties, we have performed a study on melt-spun samples of the pure and Ti or Nb substituted Laves phase $YCo_2$.

**Experiment and *ab initio* calculation methods**

The nanocrystalline $Y_{1-x}(Ti,Nb)_xCo_2$ (x = 0, 0.1) alloys were synthesized by the melt-spinning method. The initial ingots of $YCo_2$, $Y_{0.9}Ti_{0.1}Co_2$ and $Y_{0.9}Nb_{0.1}Co_2$ were prepared in an arc-furnace by repeated melting of pure Y (99.9%), Co (99.9%), Ti (99.9%) and Nb (99.9%) in a protective argon atmosphere. Subsequently, the ingots were rapidly quenched in a melt-spinning device on a rotating copper wheel with the surface velocity of 40 m/s. The X-ray diffraction (XRD) with Co-$K_\alpha$ radiation in Bragg-Brentano geometry was used to characterize the crystalline structure of produced samples. Structural parameters were determined by the Rietveld method. The mean grain size *D* was calculated from main XRD peaks between 30 and 120 degrees using the Scherrer formula. Magnetization *vs.* temperature *M(T)* and *vs.* magnetic field *M(H)* curves were measured in temperature range between 2 and 300 K with Quantum Design Physical Property Measurement System (PPMS).

In order to study electronic structure of the $YCo_2$, $Y_{0.9}T_{0.1}Co_2$, and $Y_{0.9}Nb_{0.1}Co_2$ alloys density functional theory (DFT) calculations have been performed with the full-potential local-orbital (FPLO) code.[9,10] The scalar-relativistic mode was employed in the calculations. Alloyed structures were treated by the coherent potential approximation (CPA) to take into account chemical disorder introduced by Ti and Nb substitution. The calculations were carried out for the cubic $MgCu_2$-type structure with *Fd-3m* space group (no. 227) and experimental values of the lattice constants. For the calculations we assumed the following configurations of electrons in atoms: core plus semi core (*4s4p*) plus valence (*5s5p4d*) electrons for Y and Nb atoms, and core plus semi core (*3s3p*) plus valence (*4s4p3d*) electrons for Co and Ti atoms. The calculations were performed for the reciprocal space mesh containing 1661 points within the irreducible wedge (1/48) of the Brillouin zone using the tetrahedron method[11] for integrations. The exchange-correlation potential was taken in the form proposed by Perdew and Wang.[12] Accuracies for the self-consistent calculations were set to $10^{-8}$ Ha for the total energy and $10^{-6}$ electrons for the charge.

**Results and discussion**

X-ray diffraction patterns (Fig. 1) show that obtained samples consist of a single phase which forms through polymorphous crystallization. This phase has the same stoichiometry as the initial liquid state. The structure is identified as the C15 Laves phase with $MgCu_2$-type structure and *Fd-3m* space group. In the XRD patterns of all examined alloys there are neither an amorphous "halo", nor traces of crystalline $YCo_3$ or other ferromagnetic phases (with more than 75 at.% of Co). The lattice determined for $YCo_2$ compound is equal $a = 7.223$ Å and the cell volume is $V = 376.7$ Å$^3$. Mean crystalline grain diameter $D = 50$ nm. After annealing the sample for 24 hours at 850°C the $D$ value increases to 70 nm. There is also evidence of a Laves phase with much higher lattice constant equal 7.960 Å (marked in Fig.1 with asterisk).

For $Y_{0.9}Nb_{0.1}Co_2$ the diffraction pattern is almost unchanged in comparison with $YCo_2$. There is still the same crystalline phase, but diffraction peaks are moved slightly to the higher 2Θ angle values. Atoms of Nb, with smaller diameter, partially replace Y atoms and the lattice constant diminishes to $a = 7.213$ Å. Diffraction peaks are broader than for $YCo_2$. This broadening is connected with lower value of mean grain size equal $D = 25$ nm. There also is a small volume of additional $NbCo_2$ phase with the same structure and decreased lattice constant (peak at 52 degrees and slightly increased intensity of the peak at about 44 degrees, both marked with black circles). The $Y_{0.9}Ti_{0.1}Co_2$ compound possess the same crystalline phase, but its lattice constant decreases (comparing with $YCo_2$) to $a = 7.192$ Å. The mean grain diameter is $D = 33$ nm.

Magnetism of $Y_{1-x}$(Nb or Ti)$_x Co_2$ (x = 0, 0.1) was examined to reveal the influence of topological and chemical changes on the possible appearance of ferromagnetism. The thermal variation of magnetization (zero field-cooled and field-cooled measurements) in $YCo_2$, $Y_{0.9}Nb_{0.1}Co_2$ and $Y_{0.9}Ti_{0.1}Co_2$ is plotted in Figure 2. The presence of maxima in the temperature dependences of magnetization in constant field (DC susceptibility) is evidenced for all examined compositions. Such a behavior is typical for exchange-enhanced paramagnets.[13]

All examined samples show evidence of a magnetically ordered phase, observed by an increase of magnetization at low temperatures $T < 25$ K. Both modified compounds reveal also a second magnetic anomaly by an excess of the magnetization at characteristic temperatures $T_{max} = 110$ K for $Y_{0.9}Nb_{0.1}Co_2$ and $T_{max} = 100$ K for $Y_{0.9}Ti_{0.1}Co_2$, respectively, compared to a wide maximum $T_{max} \approx 160$ K in $YCo_2$. But, the overall magnetization is enhanced in the alloyed samples.

Isothermal magnetization curves at low magnetic fields, showed in Figure 3, are comparable with curves obtained for nanocrystalline, mechanically milled $YCo_2$ [2], but there is no evidence of saturation in fields up to 7 T (inset of Fig. 3). With increasing temperature below $T_{max}$ the magnetization of the $YCo_2$ sample raises. This kind of behavior is expected but further investigations in higher magnetic fields (using pulsed fields up to 80 T) should be performed because the nanocrystalline $YCo_2$ is expected to display a metamagnetic transition similar to the IEM found in the polycrystalline compound.[14]

The $M(H)$ curves are changed for $Y_{0.9}Ti_{0.1}Co_2$ and $Y_{0.9}Nb_{0.1}Co_2$ in comparison to $YCo_2$ compound. At low temperatures the superposition of two hysteresis loops can be seen. At temperatures near $T_{max}$ (below 120 K) one of the loops disappears, and at higher temperatures the *M(H)* dependence becomes linear. The evolutions of the hysteresis loops with increasing temperature for all investigated samples are shown in Figure 3. Features observed in *M(T)* curves and shape of hysteresis loops for Nb and Ti-containing samples suggest the existence of additional magnetically ordered phase which appears below 120 K for both compositions.

From the *ab-initio* calculations the heats of formation have been obtained with the following values -102.2, -51.5 and -38.5 kJ/mol for $YCo_2$, $Y_{0.9}Nb_{0.1}Co_2$ and $Y_{0.9}Ti_{0.1}Co_2$, respectively. They differ from those calculated with Miedema's model.[15] Calculated formation enthalpies $\Delta H^{form}$ are -27.61 kJ/mol for $YCo_2$, -28.75 kJ/mol for $Y_{0.9}T_{0.1}Co_2$ and -26.82 kJ/mol for $Y_{0.9}Nb_{0.1}Co_2$ for this empirical model. That discrepancy is particularly noteworthy for $YCo_2$. Partial subsitution of Y atoms by Ti and Nb leads to the chemical disorder in the ordered crystallographic structure. The chemical disorder destabilizes the ordered intermetallic structure and drives calculated heats of formation closer to the results of the Miedema's model.

The main aspect of calculations is the stability of nonmagnetic and magnetic phases. Typical bulk $YCo_2$ is a Pauli paramagnet and would be described by a nonmagnetic solution in the calculations. Direct comparison of total energies in para- and ferromagnetic phases confirms higher stability of the nonmagnetic system by about 2.2 meV/f.u. Similar result are obtained for $Y_{0.9}Nb_{0.1}Co_2$ system, where the total energy of the nonmagnetic state is lower by about 5.5 meV/f.u. The opposite behavior is found for the $Y_{0.9}Ti_{0.1}Co_2$ alloy where magnetic order decreases the total energy by about 47 meV/f.u. compared to the nonmagnetic state.

In all cases the valence bands are dominated by *d*-electrons forming bands between -3 eV and the Fermi level. The contribution of the *s*-type electrons is significant below -3 eV to the bottom of the valence band. The chemical disorder introduced by Ti and Nb atoms results in broadened features of the DOS. The Nb and Ti substitutions introduce additional electrons to the valence band, so that the Fermi level is shifted to higher energies.

In the spin-polarized calculations for $Y_{0.9}Ti_{0.1}Co_2$, the band splitting leads to a spin polarization equal to 73%. The local magnetic moments are equal to 1.1, -0.3, and -0.9 $\mu_B$/atom for Co, Y, and Ti, respectively. The values of DOS at the Fermi level give the Sommerfeld coefficients in the linear term in the specific heat equal to 6.1, 6.4, and 9.1 mJ/(mol $K^2$) for $YCo_2$, $Y_{0.9}Nb_{0.1}Co_2$ and $Y_{0.9}Ti_{0.1}Co_2$ (FM), respectively.

Some calculations were also made for $YCo_2$ with C15-Laves phase structure, but with lattice constant $a$ = 7.960 Å. This phase is observed in X-ray diffraction pattern. The magnetic state with this lattice constant is more stable than nonmagnetic. Antiparallel magnetic moments with values 1.64 and -0.47 $\mu_B$/atom appear on Co and Y, respectively, and the total magnetic moment is about 2.80 $\mu_B$/f.u. The calculated properties of this structure suggest that the magnetization increase at low temperatures in all examined samples relies on this expanded C15-structure of $YCo_2$.

Another origin of magnetic interactions in investigated alloys could be the presence of voids or vacancies. *Ab-initio* results obtained for a $Y_7Co_{16}$ supercell based on calculations using WIEN2k code[16] showed higher stability of the magnetic solutions for systems where one Y or Co atom is removed (see Fig. 4). The stability was determined by comparing the total energies calculated for para- and ferromagnetic phases ($E_{PM}-E_{FM}$=0.1 eV/f.u.). The reduction of symmetry results in the presence of two types of Co atoms having magnetic moments equal to 1.13 and 1.44 $\mu_B$/atom. The moments on Y atoms are equal to -0.18 and -0.19 $\mu_B$/atom with opposite direction to the Co moments.

**Conclusions**

All investigated compounds show a broad maximum of the magnetization *versus* temperature *M(T)* curves similar to polycrystalline YCo$_2$ behaving as exchange-enhanced Pauli paramagnets. But there is clear increase of magnetization at lower temperatures and hysteresis loops are visible in *M(H)* curves. These characteristics are evidence for a soft ferromagnetic behavior with small coercive fields. By replacing of Y by Ti or Nb, the overall magnetization in low magnetic field increases about two times. This feature along with observed changes in *M(H)* curves are due to the appearance a magnetically ordered phase. Its origin could be connected with point defects and / or an expanded lattice structure of the YCo$_2$ phase. Any ferromagnetic impurity phase was not observed in XRD patterns. The results of theoretical calculations for Y$_{0.9}$Ti$_{0.1}$Co$_2$ show that magnetic moments on Co and Ti are oriented antiparallel with values of 1.1 $\mu_B$/atom and -0.9 $\mu_B$/atom, respectively. When taking into account magnetic moment of -0.3 $\mu_B$/atom induced on Y it gives total magnetic moment equal to 1.84 $\mu_B$/f.u. For Y$_{0.9}$Nb$_{0.1}$Co$_2$ system, the total energy for nonmagnetic state is lower by about 5.5 meV/f.u compared to the spin-polarized state. Hence, the substitution effect alone does not explain the enhanced magnetic ordering in both alloys, which rather appears to be driven by an increased density of lattice defects and decreased grain sizes.

**Acknowledgments**

Work supported by the National Science Centre within the research project no. N202 381740.

Figure Captions

Fig. 1 X-ray diffraction patterns of the melt-spun $YCo_2$, $Y_{0.9}Nb_{0.1}Co_2$ and $Y_{0.9}Ti_{0.1}Co_2$ compounds with $MgCu_2$-type C15 Laves phase. The peak marked with asterisk is due to a Laves phase structure with much higher lattice constant. A low content of an impurity $NbCo_2$ phase is marked with black circles.

Fig. 2 Temperature dependence of the magnetization for melt-spun $YCo_2$, $Y_{0.9}Ti_{0.1}Co_2$ and $Y_{0.9}Nb_{0.1}Co_2$, measured in magnetic field of 0.5 T.

Fig. 3 $YCo_2$, $Y_{0.9}Ti_{0.1}Co_2$ and $Y_{0.9}Nb_{0.1}Co_2$ low-field hysteresis loops measured at temperatures 2, 35 and 120 K. First quadrant of hysteresis loop measured for $YCo_2$ at 120 K in high magnetic fields (up to 7 T) is shown in the inset.

Fig. 4 The supercell of the $Y_7Co_{16}$ system with one vacancy at the Y site. There are two different Co-sites with spin magnetic moments of 1.13 (blue) and 1.44 $\mu_B$ (red), respectively.

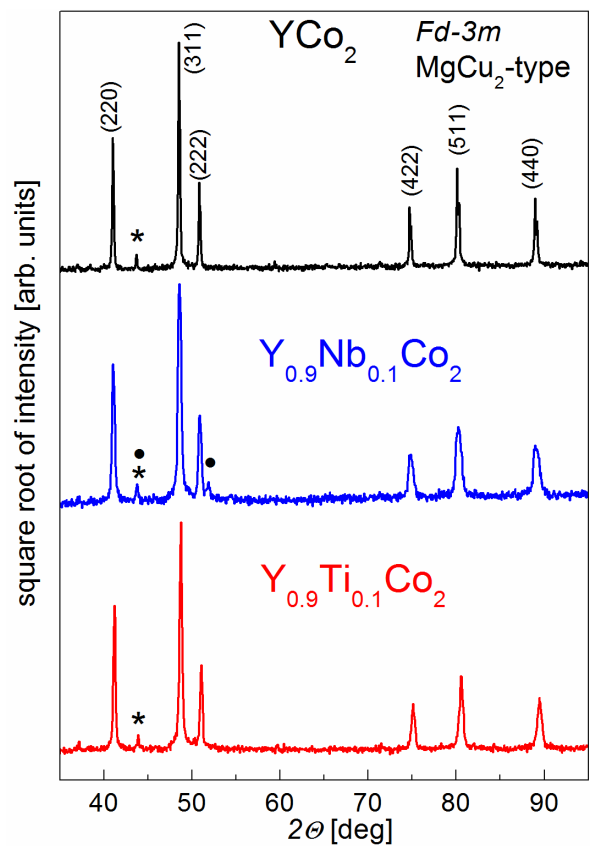

**Fig. 1.**

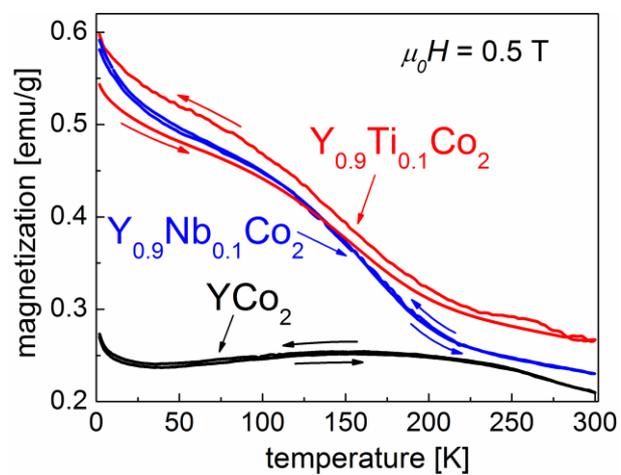

**Fig. 2.**

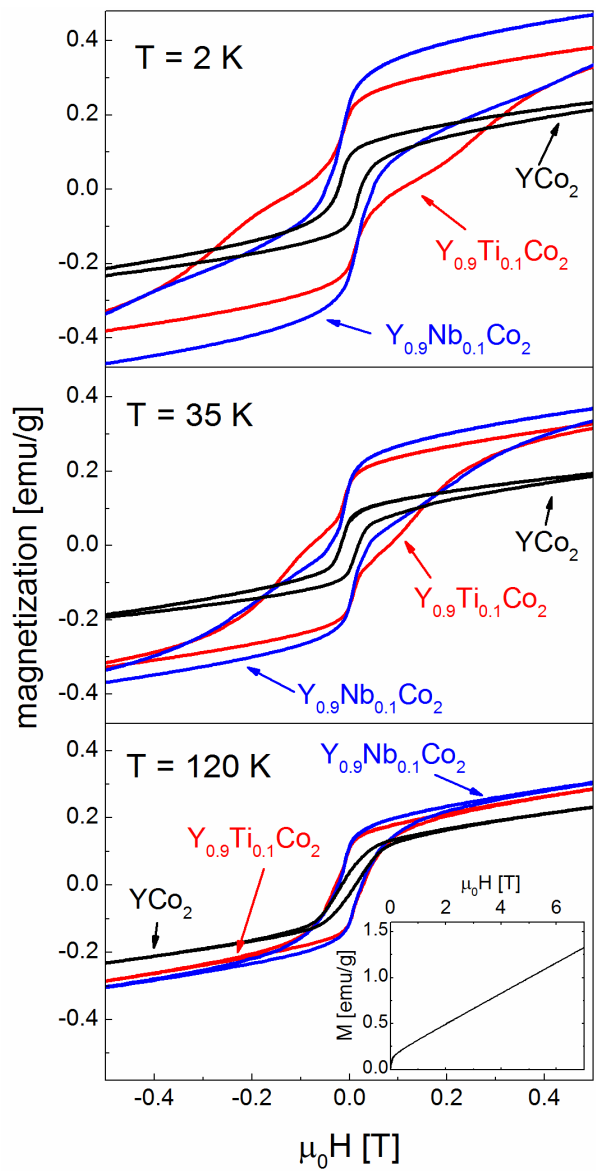

**Fig. 3.**

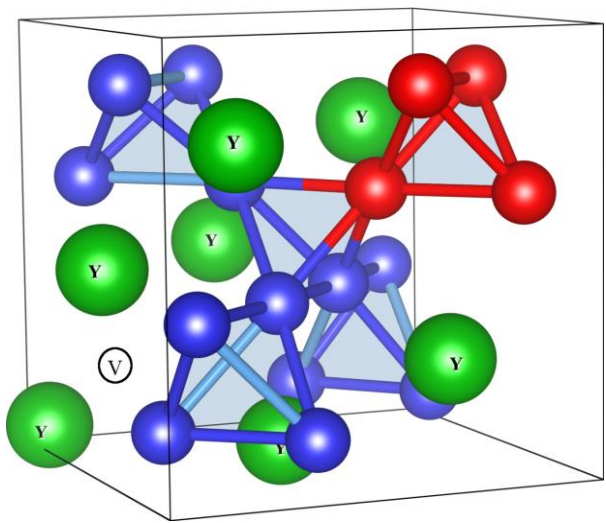

**Fig. 4.**